\documentclass{article}

\usepackage{PRIMEarxiv}

\usepackage[utf8]{inputenc} % allow utf-8 input
\usepackage[T1]{fontenc}    % use 8-bit T1 fonts
\usepackage{hyperref}       % hyperlinks
\usepackage{url}            % simple URL typesetting
\usepackage{booktabs}       % professional-quality tables
\usepackage{amsfonts}       % blackboard math symbols
\usepackage{nicefrac}       % compact symbols for 1/2, etc.
\usepackage{microtype}      % microtypography
\usepackage{lipsum}
\usepackage{fancyhdr}       % header
\usepackage{graphicx}       % graphics
\graphicspath{{media/}}     % organize your images and other figures under media/ folder

\title{Clustered Patch Embeddings for Permutation-Invariant Classification of Whole Slide Images
}

\author{
  Ravi Kant Gupta, Shounak Das, Amit Sethi \\
  Department of Electrical Engineering \\
  Indian Institute of Technology Bombay \\
  Mumbai, India\\
  \texttt{\{ravigupta131, 21D070068, asethi\}@iitb.ac.in} \\
}

\begin{document}
\maketitle

\begin{abstract}
Whole Slide Imaging (WSI) is a cornerstone of digital pathology, offering detailed insights critical for diagnosis and research. Yet, the gigapixel size of WSIs imposes significant computational challenges, limiting their practical utility. Our novel approach addresses these challenges by leveraging various encoders for intelligent data reduction and employing a different classification model to ensure robust, permutation-invariant representations of WSIs. A key innovation of our method is the ability to distill the complex information of an entire WSI into a single vector, effectively capturing the essential features needed for accurate analysis. This approach significantly enhances the computational efficiency of WSI analysis, enabling more accurate pathological assessments without the need for extensive computational resources. This breakthrough equips us with the capability to effectively address the challenges posed by large image resolutions in whole-slide imaging, paving the way for more scalable and effective utilization of WSIs in medical diagnostics and research, marking a significant advancement in the field.
\end{abstract}

% keywords can be removed

\section{Introduction}
In the evolving field of digital pathology, Whole Slide Imaging (WSI) has emerged as a transformative technology, enabling the digitization of histopathological slides at gigapixel resolution. This advancement has not only facilitated remote diagnostics and educational opportunities but also opened new avenues for quantitative image analysis~\cite{pantanowitz2011review, malarkey2015utilizing}. Despite its potential, the sheer size and complexity of WSIs pose significant computational challenges, limiting the practicality of large-scale analysis and the application of advanced machine learning techniques~\cite{brachtel2012digital, kumar2020whole}. Whole slide imaging (WSI) represents a significant breakthrough in digital pathology, enabling the digitization of histological slides at high resolutions. This advancement allows for improved visualization, analysis, and management of tissue samples, essential for accurate disease diagnosis and research. However, the sheer size and complexity of WSIs pose unique challenges in image processing and analysis, necessitating innovative approaches for efficient and effective feature extraction and classification.

Traditional methods for analyzing WSIs often rely on supervised learning techniques, which require extensive annotated datasets prepared by expert pathologists. This process is not only time-consuming but also prone to variability due to inter-observer differences. Moreover, the high-dimensional nature of WSIs results in computational and storage challenges, limiting the scalability of conventional approaches. In response to these challenges, recent research has explored the potential of self-supervised learning~\cite{chen2020simple,li2021dual} as a promising alternative for feature extraction from WSIs. Self-supervised learning, a subset of unsupervised learning methods, involves generating labels from the data and using them as supervisory signals for learning rich, discriminative features. This technique is particularly advantageous in the context of WSIs, where labeled data are scarce and costly to obtain.

Our research introduces a comprehensive framework that leverages various encoders to preprocess WSIs and extract meaningful features. This novel approach begins with an advanced preprocessing stage designed to enhance the visual quality of WSIs and prepare them for subsequent analysis. We have used a deep learning model~\cite{PATIL2023100306} to get the WSI without artifacts. After that, we employ adaptive filtering techniques to improve contrast and highlight pertinent morphological details, which are essential for effective feature learning. A preprocessed WSI is divided into patches of size 512x512 to manage the high resolution of the image. Following preprocessing, we implement SimCLR~\cite{chen2020simple}, ResNet50~\cite{jian2016deep}, EfficientNet~\cite{tan2019efficientnet}, RegNet~\cite{7995968}, ConvNeXT\_Tiny~\cite{liu2022convnet} and Swin\_Tiny~\cite{liu2021swin} model to derive a high-dimensional feature space from the enhanced images. 

% This model is trained using a contrastive loss~\cite{chen2020simple} mechanism, which encourages the learning of features that are invariant to minor perturbations in the input data, thus capturing the intrinsic patterns and textures characteristic of different histological tissues.

To address the challenge of high dimensionality in the learned feature space, we introduce an innovative clustering~\cite{sharma2021cluster} approach. Instead of using all extracted features, our method focuses on clustering these features and uses the centroids of these clusters as new, compact representations of the original WSIs. This step significantly reduces the dimensionality of the data, making it more manageable for subsequent analysis and classification tasks. The clustered mean vectors serve as the input to our classification module, which are Swin\_Tiny, Multi-layer perceptron(MLP)\cite{haykin1994neural} and attention-based multiple instance learning (MIL)~\cite{maron1997framework} to mitigate permute variance problem of mean feature vector obtained from clustering. Recent advances in transformer models have revolutionized natural language processing and are beginning to show significant potential in image-based tasks~\cite{han2022survey, wu2020visual}. These models, known for their ability to handle sequence data, offer an intriguing solution to the problem of permutation invariance in image patches derived from WSIs, a challenge that needs to be adequately addressed by conventional convolutional neural networks (CNNs). Due to these features of the transformer, we have probed both the techniques, such as MIL and Transformer, as classifiers. Transformer models, known for their effectiveness in handling sequential data, are adapted in our framework to process the spatial relationships among cluster centroids. These models use self-attention mechanisms to weigh the importance of different regions in a WSI, allowing the classifier to focus on areas most indicative of the pathological state.

Additionally, we investigate with MLP \cite{haykin1994neural} and attention-based MIL~\cite{ilse2018attention} component to further refine our classification strategy. MIL is particularly well-suited for tasks where labels are available at a coarse level (e.g., slide level) rather than a fine-grained level (e.g., pixel level). In our approach, each WSI is treated as a "bag" of instances (cluster centroids), and the MIL classifier learns to attend to those instances most relevant for predicting the slide's label. The use of transformers and attention-based MIL as a classifier not only enhances the classification accuracy but also improves the model's interpretability. By focusing on specific clusters within a WSI, pathologists can identify the critical areas that led to a particular diagnostic decision, thereby aligning the model's operation with clinical reasoning processes.

\begin{figure*} 
\centering
\includegraphics[height=7cm,width=12.5cm]{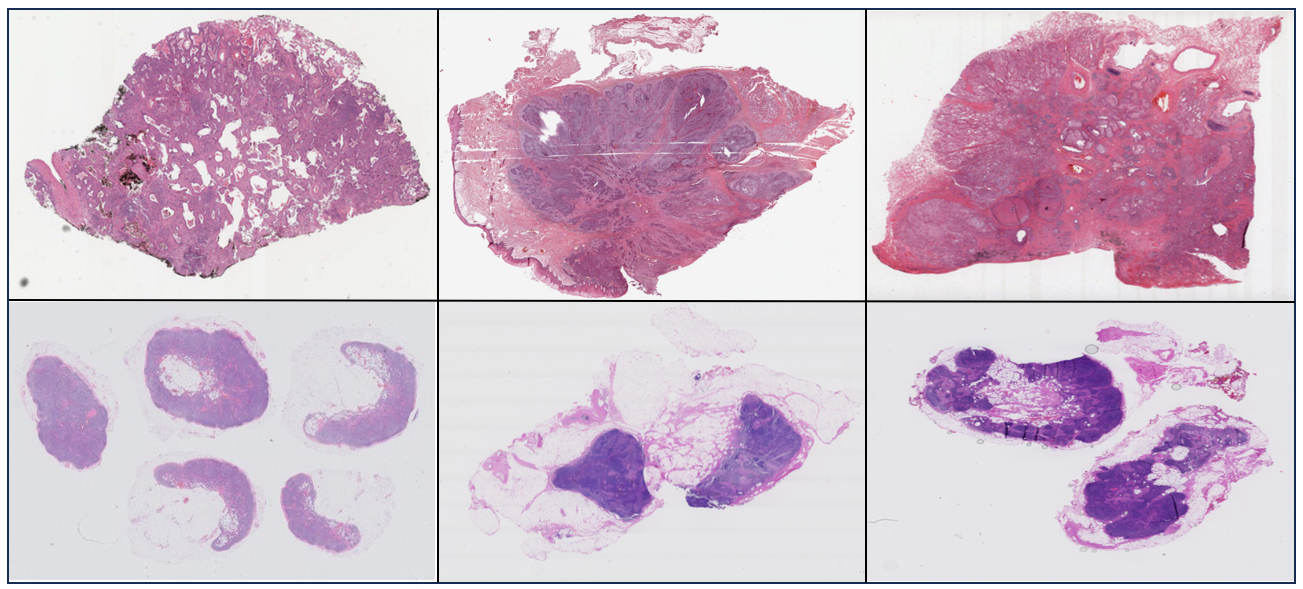}
\vspace{0.25cm}
\caption{Thumbnail image samples of TCGA Lung data~\cite{coudray2018classification} (first row) and Camelyon17 dataset~\cite{bandi2018detection} (bottom row)}
\label{fig1}
\end{figure*}

To validate the effectiveness of our proposed methodology, we conducted extensive experiments on a diverse dataset of WSIs such as Lung Cancer~\cite{coudray2018classification,gupta2023egfr} and Camelyon17~\cite{bandi2018detection} dataset. The snapshot of thumbnails of TCGA LUNG data and Camelyon17 dataset is shown in Fig.~\ref{fig1}. Our results demonstrate significant improvements in classification accuracy compared to traditional methods, highlighting the potential of our approach to transform the landscape of digital pathology.
In conclusion, our research contributes to the fields of digital pathology and machine learning by presenting a novel, effective, and efficient framework for the preprocessing, feature extraction, and classification of WSIs. This approach not only addresses the limitations of existing techniques but also paves the way for more scalable and insightful analyses of histological data, with potential applications in automated disease diagnosis and biomarker discovery.

\section{Related Work}
Learning or extracting robust and discriminative features is vital for image classification. These features can be of two types -- handcrafted features and features learned from the data itself. Typically, classification methods for whole slide images (WSI) are grouped based on various factors such as their pooling techniques, the assumptions underlying their models, or the specific challenges they aim to tackle~\cite{dimitriou2019deep,ibrahim2020artificial}. 

One of the main challenges in WSI processing is their size. A typical WSI may have more than 100,000 pixels in each direction. Because images of this large a size cannot be processed in one go by a convolutional neural network (CNN)~\cite{albawi2017understanding} or a vision transformer (ViT)~\cite{khan2022transformers}, the WSI is broken into patches (sub-images) of a more manageable size. In a multiple instance learning (MIL)~\cite{maron1997framework} framework, each WSI is typically conceptualized as a `bag.' In this context, an instance might be a randomly selected patch from the WSI with or without its corresponding (i, j) coordinates within the WSI. Variations of the standard MIL assumption have been effectively applied in WSI classification~\cite{hou2016patch,combalia2018monte,li2019refinenet,chen2019rectified}. Taking inspiration from the embedding-based methods in multiple instance learning, a common practice is to extract a set of patches, such as $\{x_1, ..., x_n\}$, from a WSI. These patches are then processed through a convolutional neural network (CNN), denoted as f, resulting in a set of vectors $\{f(x_1), ..., f(x_n)\}$. CNN-based \cite{albawi2017understanding} and Transformer-based~\cite{khan2022transformers} are the major backbone to obtain meaningful features. But nowadays, SSL~\cite{chen2020simple,li2021dual} is being used by many researchers. These vectors are subsequently combined into a single vector, which essentially represents the encoded information of the WSI from which the patches $\{x_1, ..., x_n\}$ are derived. For combining these values, clustering~\cite{sharma2021cluster} strategies have been suggested. Research indicates that the selection of clustering strategy and the number of clusters significantly impacts performance, both for natural ~\cite{omran2005dynamic} and histopathological images ~\cite{sharma2021cluster}. A method closely related to ours in terms of data preprocessing, is found in the study of ~\cite{couture2018multiple} and ~\cite{akbarnejad2021deep}, where they ~\cite{couture2018multiple} process a large image by calculating various quantiles of $f(x)$, with $x$ being a randomly selected patch.  And ~\cite{akbarnejad2021deep} designed a special purpose dataloader to get the patches from whole slide images. For WSI encoding, ~\cite{huang2019convolutional} suggests using a generative model, such as a variational auto-encoder or BiGAN, trained on patches from WSIs. Each WSI is then divided into a patch grid, with each patch (or grid cell) fed into the encoder of the trained generative model. Mean cluster representation for feature extraction from these patches is known for robustness in capturing fine-grained image details. To address the permute invariance problem of the mean vector obtained from clustering is addressed by Attention-MIL and  Transformer models. These models combine information from multiple regions of a large whole slide image. In our method, we have tried to find the alternate approach for Fisher Vector~\cite{akbarnejad2021deep}.

\section{Methodology}

\begin{figure*} 
\centering
\includegraphics[height=4cm,width=16cm]{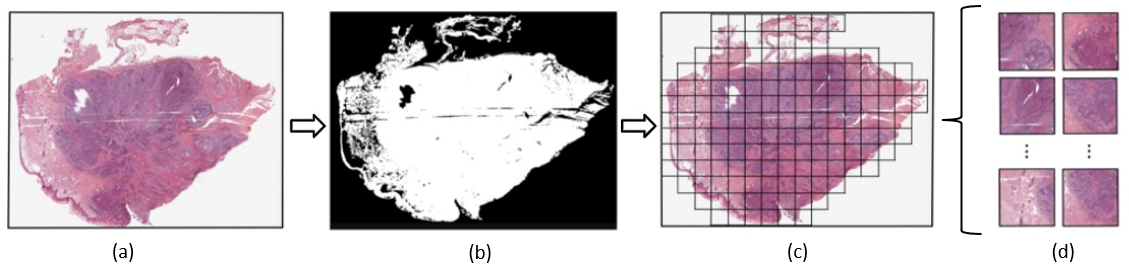}
\vspace{0.25cm}
\caption{Preprocessing pipeline: (a) Snapshot of sample image, (b) Tissue detection, (c) Tiling of WSI, and (d) Patches}
\label{preprocessing}
\end{figure*}

In our study, we developed an efficient approach for analyzing whole slide images (WSIs). Initially, we utilized a deep learning model~\cite{PATIL2023100306} to remove artifacts from the WSIs. Following this, we applied an automated tissue detection algorithm that identifies regions of interest (ROIs) through color thresholding and morphological operations, leveraging the capabilities of HistomicsTK~\cite{HistomicsTK}. Our proposed method is mainly centered on getting a set of features as minimum as possible to represent the WSI and having the capability to describe the image in terms of classification task. To do so, we performed a
set of preprocessing which includes, tissue detection, patching, quality patches filtering, and nucleus count. We excluded the patches having nuclei count$<$10 as the Region of Interest is the tumor region. To perform tissue detection, we have used a Python library HistomicsTK~\cite{HistomicsTK}
(shown in Fig. 2(b)). After obtaining the tissue, we performed patching using a sliding window algorithm (shown in
Fig. 2(c)). While applying the sliding window algorithm
to get the patches, we extended this algorithm to filter out
noisy patches. We extracted non-overlapping patches of size 512x512 pixels at
40x zoom level using the OpenSlide library. For clean patches
obtained from the previous process, we performed a nucleus
count using the library HistomicsTK for every patch to get
the patches with high nuclei count. The core idea behind
this is to get a more informative region of WSI (shown in
Fig. 2(d)). This critical first step ensures that only tissue areas with diagnostic value are selected, filtering out large sections of white space and non-diagnostic material prevalent in WSIs. Following identifying ROIs, we extracted 512x512 pixel patches shown in Fig.~\ref{preprocessing}, balancing the need to preserve histological detail with computational efficiency while minimizing overlap to ensure a diverse representation of the tissue features.

\begin{figure*} 
\centering
\includegraphics[height=5cm,width=14cm]{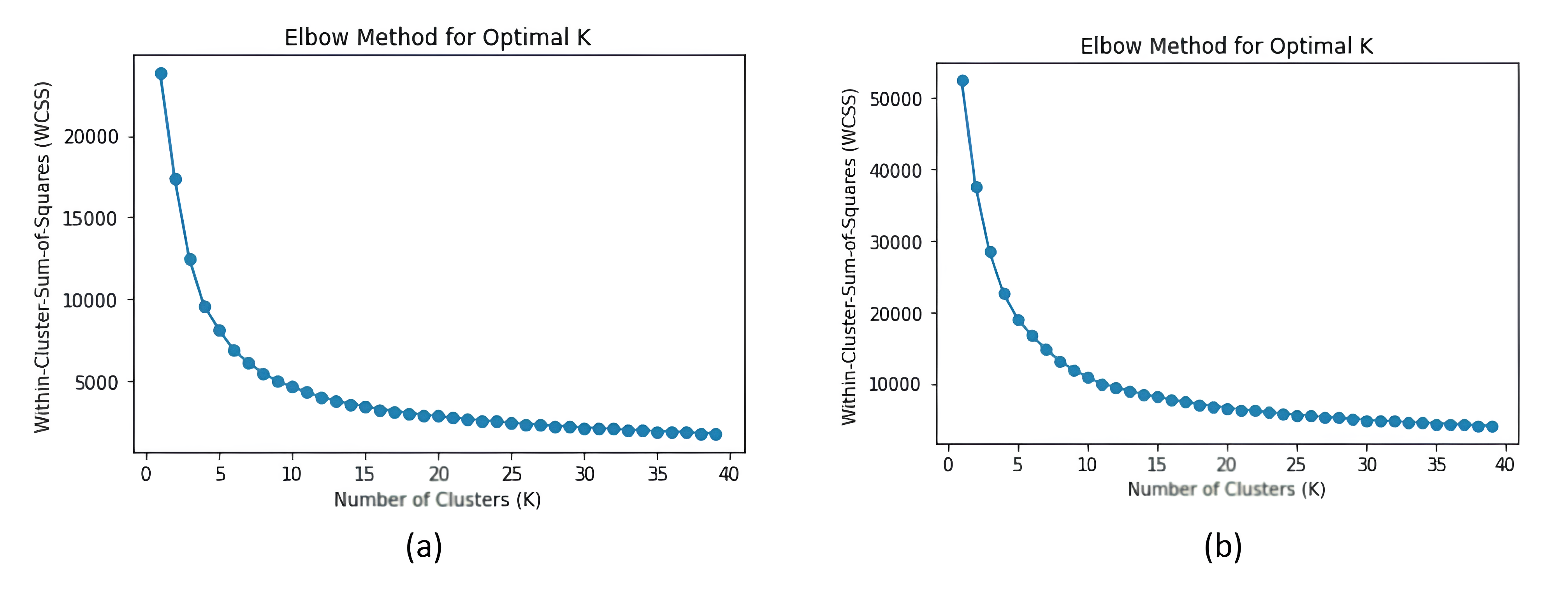}
\caption{Elbow plot illustrating the variation of within-cluster sum of squares (WCSS) for k = 2 to 30 in K-means clustering using the (a) TCGA Dataset and (b) Camelyon17 Dataset}
\label{elbow}
\end{figure*}

\begin{figure*} 
\centering
\includegraphics[height=9cm,width=14cm]{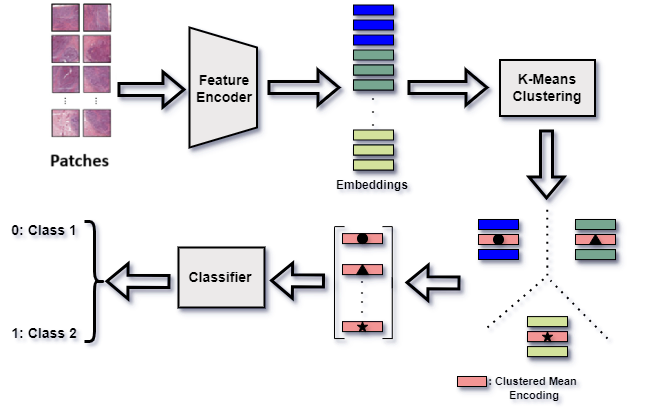}
\caption{Model Workflow}
\label{fig:main_pic}
\end{figure*}

To distill meaningful features from these patches, we utilized various pre-trained encoders such as SimCLR~\cite{chen2020simple}, ResNet50~\cite{jian2016deep}, EfficientNet~\cite{tan2019efficientnet}, RegNet~\cite{7995968}, ConvNeXT\_Tiny~\cite{liu2022convnet} and Swin\_Tiny~\cite{liu2021swin}, to transforms raw pixel data into high-dimensional feature vectors. These vectors were then clustered using the K-means algorithm~\cite{lloyd1982least,macqueen1967some}, where the number of clusters was empirically determined using Elbow Method~\cite{syakur2018integration} to achieve an optimal balance between computational efficiency and the preservation of pathological diversity. The Elbow Method determines the optimal number of clusters in K-means by identifying where the within-cluster sum of squares (WCSS) begins to decrease linearly, indicating a diminishing return on adding more clusters. Fig.~\ref{elbow} shows the optimal number of clusters. For our dataset, we find an optimal number of clusters as 10. The mean feature vector of each cluster was computed to represent the collective characteristics of the patches within, and these mean vectors were concatenated to form a single, comprehensive feature vector for each WSI. After concatenation, the major challenge was how we can make it permute invariant. To mitigate this challenge, we experimented with various permute invariant classifiers such as Swin-Tiny, MLP, and attention MIL. Hence, Classification was performed using a fine-tuned Swin-Transformer model~\cite{liu2021swin}, MLP \cite{haykin1994neural} and Attention MIL~\cite{ilse2018attention}, capable of handling the permutation-invariant feature vectors and ensuring that the classification process was based on the histopathology content rather than the order of feature extraction. This methodology not only streamlines the analysis of WSIs but also enhances the computational efficiency and accuracy of pathological assessments relatively. By getting attention to the mean of clusters also helps in highlighting regions of interest in a huge WSI to probe the investigation for pathologists. The workflow as a whole as shown in Fig.~\ref{fig:main_pic} depicts a clear idea of our methodology.

% \begin{figure*}
%     \centering
%     \includegraphics[width=0.4\textwidth]{tcga.jpg} % Replace 'tcga.jpg' with the filename of your first image
%     \hspace{0.2cm} % Adjust the horizontal space between the two images
%     \includegraphics[width=0.4\textwidth]{camelyon.jpg} % Replace 'camelyon.jpg' with the filename of your second image
%     \caption{Caption for both images}
%     \label{fig:both_images}
% \end{figure*}

\begin{table*}[ht!]
\begin{center}
\begin{tabular}{|c|c|c|c|c|c|c|}
\hline
 Feature Extractor	 &	Clustering &	Classifier	&Accuracy & Precision & Recall	\\
\hline
 ResNet-50\cite{jian2016deep}	 & No	&	MLP \cite{haykin1994neural}	&	0.71&	0.71&	0.74\\
	
\hline
 ResNet-50 \cite{jian2016deep} & Yes	&	AMIL \cite{ilse2018attention}	&	\textbf{0.83}& 0.81	& 0.83	\\
\hline
 ResNet-50 \cite{jian2016deep} & Yes	&	Swin-Transformer \cite{liu2021swin}	&	0.73&	0.74&	0.71\\
\hline
 SimCLR \cite{chen2020simple} & Yes	&	Swin-Transformer \cite{liu2021swin}&	0.75&	0.74&	0.78\\
\hline
 SimCLR \cite{chen2020simple} & Yes	&	AMIL \cite{ilse2018attention} &	{0.81}&	{0.80}&	\textbf{0.84}\\
 \hline
 Swin Tiny \cite{liu2021swin} & No	&	AMIL \cite{ilse2018attention} &	{0.75}&	{0.75}&	{0.75}\\
 \hline
 Swin Tiny \cite{liu2021swin} & Yes	&	AMIL \cite{ilse2018attention}&	{0.67}&	{0.70}&	{0.62}\\
 \hline
  ConvNeXT \cite{liu2022convnet} & Yes	&	MLP \cite{haykin1994neural}&	{0.72}&	{0.70}&	{0.74}\\
 \hline
 ConvNeXT \cite{liu2022convnet} & Yes	&	Swin-Transformer \cite{liu2021swin} &	{0.63}&	{0.63}&	{0.63}\\
 \hline
 ConvNeXT \cite{liu2022convnet} & No	&	AMIL \cite{ilse2018attention}&	{0.72}&	{0.75}&	{0.70}\\
 \hline
 ConvNeXT \cite{liu2022convnet} & Yes	&	AMIL \cite{ilse2018attention}&	{0.81}&	\textbf{0.81}&	{0.75}\\
 \hline
   EfficientNet \cite{tan2019efficientnet}  & Yes	&	MLP \cite{haykin1994neural}&	{0.69}&	{0.66}&	{0.70}\\
 \hline
 EfficientNet \cite{tan2019efficientnet}  & Yes	&	Swin-Transformer \cite{liu2021swin} &	{0.72}&	{0.67}&	{0.72}\\
 \hline
 EfficientNet \cite{tan2019efficientnet} & No	&	AMIL \cite{ilse2018attention}&	{0.67}&	{0.70}&	{0.62}\\
 \hline
 EfficientNet \cite{tan2019efficientnet} & Yes	&	AMIL \cite{ilse2018attention}&	{0.75}&	{0.75}&	{0.75}\\
  \hline
  RegNet \cite{7995968} & Yes	&	MLP \cite{haykin1994neural}&	{0.75}&	{0.72}&	{0.75}\\
 \hline
 RegNet \cite{7995968} & Yes	&	Swin-Transformer \cite{liu2022convnet}&	{0.67}&	{0.72}&	{0.6}\\
 \hline
 RegNet \cite{7995968} & No	&	AMIL \cite{ilse2018attention}&	{0.66}&	{0.70}&{0.63}\\
 \hline
 RegNet \cite{7995968} & Yes	&	AMIL \cite{ilse2018attention}&	{0.75}&	{0.75}&	{0.75}\\
  \hline

\end{tabular}
\end{center}  
\caption{ Results (\%) on the Lung dataset for binary classification. The best performance is marked as bold.}
\label{table1}
\end{table*}

\begin{table*}[ht!]
\begin{center}
\begin{tabular}{|c|c|c|c|c|c|c|c|}
\hline
 Backbone Feature Extractor	 &	Clustering &	Classifier	&Accuracy & Kappa Score & Precision & Recall	\\
\hline
 ResNet-50	\cite{jian2016deep} & No	&	AMIL \cite{ilse2018attention}	&	0.72& 0.54&	0.73&	0.72\\
	
\hline
 ResNet-50 \cite{jian2016deep}  & Yes	&	AMIL	\cite{ilse2018attention}&	\textbf{0.75}&0.56	& \textbf{0.80}&	\textbf{0.75}\\
\hline
 SimCLR \cite{chen2020simple} & No	&	AMIL \cite{ilse2018attention}	&	0.64&	0.48 & 0.64&	0.66\\
\hline
 SimCLR \cite{chen2020simple} & Yes	&	AMIL \cite{ilse2018attention}	&	0.68& 0.50 &	0.70&	0.68\\
\hline
 ConvNeXT \cite{liu2022convnet} & No	&	AMIL \cite{ilse2018attention}	&	0.58& 	0.42 & 0.60&	0.60\\
\hline
 ConvNeXT \cite{liu2022convnet} & Yes	&	AMIL	\cite{ilse2018attention}&	0.72& \textbf{0.57} &	0.72&	0.75\\
\hline
\end{tabular}
\end{center}  
\caption{ Results (\%) on the Camelyon17 dataset for binary classification. The best performance is marked as bold.}
\label{table2}
\end{table*}

\section{Experimentation and Dataset}
\subsubsection{\textbf{Experimentation}}
% In our study, we utilized a modified ResNet50~\cite{jian2016deep} architecture and MobileNetV3small~\cite{howard2019searching} architecture trained on ImageNet~\cite{russakovsky2015imagenet} as our base model, where we removed the final pooling and fully connected layers. Subsequently, we integrated a 1x1 convolutional layer to compress the channel count, effectively reducing the descriptor space dimensionality to 10. This was followed by the addition of an instance normalization layer configured for 10 channels. For our experiments, we set the number of centers for Fisher vector coding (denoted as m in the equation) to five. Adhering to the notation in \cite{arun2020enhanced}, we configured the Fisher vector encoding parameters with $\pi_{m}$ at 0.2 and $\sigma_{m}$ at 0.1. An average pooling layer was then incorporated. The training process involved a batch size of 1 over 500 epochs, utilizing the AdamW optimizer with a learning rate of 0.00001 and weight decay of 0.00001.
In our study, we utilized various pre-trained encoders such as SimCLR~\cite{chen2020simple}, ResNet50~\cite{jian2016deep}, EfficientNet~\cite{tan2019efficientnet}, RegNet~\cite{7995968}, ConvNeXT\_Tiny~\cite{liu2022convnet} and Swin\_Tiny~\cite{liu2021swin}, trained on ImageNet~\cite{russakovsky2015imagenet} as our base model for feature extraction from the patches.  We then applied K-Means Clustering using the elbow method to cluster the features and reduce the dimension of the feature space. For our experiments, we set the number of clusters (k) to 10  for the features of both datasets (TCGA and Camelyon-17) based on the elbow plot ~\ref{elbow}. Then, for augmenting the clustered features, we applied feature scaling with scale in range (0.9, 1), feature jittering with a jitter level of 0.01, and feature mixup with $\alpha = 0.2$. Subsequently, we passed these augmented features to an Attention Multi-Instance Learning (AMIL) block, utilizing the AdamW optimizer with a learning rate of 0.001 and weight decay of 0.0001.

\subsubsection{\textbf{Dataset}}
The Camelyon17 dataset, central to the Camelyon17 Challenge, is a pivotal resource in digital pathology, primarily focused on the automated detection of breast cancer metastases in lymph node WSIs. Comprising high-resolution images with expert-annotated metastatic regions, it serves as a standard for training and testing AI models in cancer diagnosis. The dataset presents a challenging task due to the variability in metastatic tissue appearance, making it a cornerstone for advancements in computational pathology.  This dataset, building upon its predecessor CAMELYON16~\cite{litjens20181399}, provides a diverse and realistic representation of clinical scenarios with detailed annotations by expert pathologists. These annotations include various metastasis categories, such as macro-metastases, micro-metastases, and isolated tumor cells, making them invaluable for developing and testing machine-learning models. This dataset consists of 500 WSIs from different centers having four classes (Negative, Isolated Tumor Cell (ITC), Macro-metastases, and Micro-metastases). We planned our experiments as a binary classification of Metastasis Positive(includes ITC, Macro, and Micro) vs Negative. As we are experimenting with fewer WSIs of the Camelyon17 dataset, we need a baseline with similar settings. Additionally, our model underwent rigorous evaluation on the TCGA Lung dataset, specifically targeting variants in genetic mutations. Among these, the EGFR mutation stood out due to its aggressive nature, prompting a focused binary classification approach: EGFR mutations versus Non-EGFR mutations. This experimental design utilized a dataset comprising 159 slides, with 79 slides exhibiting EGFR mutations and the remaining 80 slides categorized under Non-EGFR mutations. This binary classification framework allows for precise discrimination between aggressive EGFR mutations and other genetic variants within the dataset.

\section{Results}
Table~\ref{table1} illustrates the outcomes of the binary classification
task on the TCGA Lung dataset, focusing on distinguishing between EGFR-positive (Mutated) and Non-EGFR (Wild-type) samples, using different pre-trained backbone feature extractors such as SimCLR~\cite{chen2020simple}, ResNet50~\cite{jian2016deep}, EfficientNet~\cite{tan2019efficientnet}, RegNet~\cite{7995968}, ConvNeXT\_Tiny~\cite{liu2022convnet} and Swin\_Tiny~\cite{liu2021swin}. These features are clustered using K-mean with 10 clusters. After clustering, we replace each cluster with a vector as cluster mean. This means vectors are input for various classifiers such as Swin-Transformer, MLP, and AMIL results comparable results with the classical approach used in ~\cite{gupta2023egfr}. Our proposed method not only reduces the requirement of memory but also finds a single vector(cluster mean vector) to catch the details of a gigapixel whole slide image. An ablation study was also conducted for comparison, wherein we have chosen different combinations of models used for the complete pipeline, and the results are listed in Table~\ref{table1}. This improvement, observed with the use of a larger number of patches in combination with the smaller model, suggests that regions of high cellularity play a significant role in determining the EGFR class. Our findings are that EGFR mutated regions are tumor areas with high cellular density, and clustering helps in finding global representation.
Table~\ref{table2} presents the classification outcomes on the Camelyon17 dataset, focusing on the binary differentiation of
metastasis-positive versus metastasis-negative samples. In this
analysis, we have used the same pipeline as used for Lung data. An ablation study was also conducted for comparison, wherein we have chosen different combinations of models used for the complete pipeline, and the results are listed in Table~\ref{table2}. These findings suggest that regions of high cellularity may not always serve as reliable morphological biomarkers for metastasis detection.

% \section{Conclusion}
\section{Conclusions and Discussions}
In this study, we developed a comprehensive framework for the preprocessing, feature extraction, clustering, and classification of whole slide histology images (WSIs). By leveraging multiple encoders for feature extraction and integrating clustering techniques, our approach effectively addresses critical challenges in digital pathology, including the high dimensionality of WSI data and the scarcity of labeled samples. To reduce model complexity with transformer-based encoders, we have used tiny versions of these. Our findings demonstrate that the use of robust preprocessing and advanced feature extractors significantly enhances the spatial representation of WSIs, thereby improving the quality of downstream analysis. The incorporation of clustering methods further reduces the dimensionality of the feature space, preserving essential histological information necessary for robust classification.

Moreover, the application of transformer, MLP, and attention-based Multiple Instance Learning (MIL) substantially boosts classification accuracy and interpretability, making it well-aligned with clinical diagnostic workflows. This framework also assists pathologists by directing focus towards the most diagnostically relevant regions within the WSI, improving efficiency in clinical practice. Our approach matches the traditional methods in terms of efficiency and accuracy across diverse WSI datasets, underscoring its potential as a scalable, automated solution for advancing personalized medicine and biomarker discovery. A key insight from our results across various datasets is that no single encoder or classifier is universally optimal; rather, the highest classification accuracy is achieved by selecting the most suitable combination of encoder and classifier for each specific dataset.
Future enhancements may include integrating genomic or clinical data to enrich features, expanding datasets for broader applicability, and advancing MIL algorithms to address WSI variability and complexity. Investigating alternative clustering techniques and optimizing transformers for spatial data could further improve performance.
%Bibliography
\bibliographystyle{unsrt}
\bibliography{references}

\end{document}